# Microphysical Prescriptions for Parameterized Water Cloud Formation on Ultra-cool Substellar Objects

James Mang[1,6] , Caroline V. Morley[1] , Tyler D. Robinson[2,3,4] , and Peter Gao[5]
[1] Department of Astronomy, University of Texas at Austin, Austin, TX 78712, USA; j_mang@utexas.edu
[2] Lunar & Planetary Laboratory, University of Arizona, Tucson, AZ 85721, USA
[3] Habitability, Atmospheres, and Biosignatures Laboratory, University of Arizona, Tucson, AZ 85721, USA
[4] NASA Nexus for Exoplanet System Science Virtual Planetary Laboratory, University of Washington, Box 351580, Seattle, WA 98195, USA
[5] Earth & Planets Laboratory, Carnegie Institution for Science, 5241 Broad Branch Road NW, Washington, DC 20015, USA
*Received 2024 April 11; revised 2024 August 5; accepted 2024 August 5; published 2024 October 11*

## Abstract

Water must condense into ice clouds in the coldest brown dwarfs and exoplanets. When they form, these icy clouds change the emergent spectra, temperature structure, and albedo of the substellar atmosphere. The properties of clouds are governed by complex microphysics but these complexities are often not captured by the simpler parameterized cloud models used in climate models or retrieval models. Here, we combine microphysical cloud modeling and 1D climate modeling to incorporate insights from microphysical models into a self-consistent, parameterized cloud model. Using the 1D Community Aerosol and Radiation Model for Atmospheres (CARMA), we generate microphysical water clouds and compare their properties with those from the widely used EddySed cloud model for a grid of Y dwarfs. We find that the mass of water condensate in our CARMA water clouds is significantly limited by available condensation nuclei; in models without additional seed particles for clouds added, the atmosphere becomes supersaturated. We incorporate water latent heat release in the convective and radiative parts of the atmosphere and find no significant impact on water-ice cloud formation for typical gas giant compositions. Our analysis reveals the CARMA cloud profiles have a gradual decrease in opacity of approximately 4% per bar below the cloud base. Incorporating this gradual cloud base falloff and a variable $f_{sed}$ parameter allows spectra generated from the parameterized EddySed model to better match those of the microphysical CARMA model. This work provides recommendations for efficiently generating microphysically informed water clouds for future models of cold substellar objects with H/He atmospheres.

*Unified Astronomy Thesaurus concepts:* Brown dwarfs (185); Y dwarfs (1827); Exoplanet atmospheres (487); Extrasolar gaseous giant planets (509); Planetary atmospheres (1244); Atmospheric clouds (2180)

## 1. Introduction

Brown dwarfs bridge the gap between stars and planets and can be used as analogs for temperate giant planets, sharing similar effective temperatures and masses (Cushing et al. 2011). The coldest spectral type of brown dwarfs are called Y dwarfs and have effective temperatures less than ~500 K. While warmer Y dwarfs are largely cloud-free, previous observations of the coldest Y dwarfs, with effective temperatures below ~400 K, show muted spectral features and photometric colors indicative of water-ice clouds present in the atmosphere (Faherty et al. 2014; Leggett et al. 2015; Skemer et al. 2016; Morley et al. 2018). These water-ice clouds may dampen observed spectral features, change the temperature structure, increase the albedo, and drive variability in brown dwarfs and cold giant planets.

### 1.1. Ongoing and Upcoming Observations of Cold Atmospheres

The James Webb Space Telescope (JWST) is observing the spectra of dozens of Y dwarfs with effective temperatures of 250–450 K within its first two years of operation. Notably, JWST is capable of observing wavelengths from 0.6 to 28 $\mu$m at spectral resolutions surpassing those of previous ground-based and space-based observations, with several programs (GO 2124 PI: Faherty, GO 2327 PI: Skemer, GO 2243 PI: Matthews, GTO 1230 PI: Alves de Oliveira) targeting cold brown dwarfs and temperate giant planets. Among them, WISE J085510.83-071442.5 (hereafter WISE 0855), with an anticipated effective temperature of approximately 250 K (Luhman 2014), stands out as the coldest known Y dwarf. This makes WISE 0855 a prime candidate for testing next-generation atmospheric models that incorporate water-ice clouds, a crucial step in preparing for future exoplanetary atmospheric studies. Initial analysis of JWST NIRSpec data from GTO 1230 did not definitively detect water clouds in its atmosphere (Luhman et al. 2024), but further insights regarding their presence may be offered by the MIRI data from this program as well as the time-series data from GO 2327.

In addition to Y dwarfs, JWST is also the first facility capable of directly imaging Jupiter analogs discovered by radial-velocity surveys like Eps Ind Ab (Feng et al. 2019), which has a predicted effective temperature of ~186 K based on evolutionary models (Saumon & Marley 2008). GO 2243 is the first program directly imaging a temperate giant exoplanet with JWST, with more objects planned (e.g., GO 3337 PI: Bardalez Gagliuffi, GO 4050 PI: Carter, Survey 4430 PI: Limbach, GO 4829 PI: Boccaletti, GO 4982 PI: Ruffio, GO 5037 PI: Matthews, GO 5229 PI: Matthews, GO 5835 PI:

Carter, and Survey 6005 PI: Biller). Beyond JWST, the launch of the Nancy Grace Roman Space Telescopes (NGRST) will enable more direct imaging of ultra-cool objects with water clouds in their atmospheres (Morley et al. 2014b; Lupu et al. 2016; Lacy et al. 2019). The Roman Coronagraph Instrument, while a technology demonstration, will potentially have the capability to observe down to Saturn-sized planets and provide images and low-resolution spectra of mature giant exoplanets similar to Jupiter. With these observational advancements, it is crucial to improve our ability to generate accurate atmospheric models of cold substellar objects hosting water clouds.

### 1.2. The Current Landscape of Cold Atmosphere Models

Currently, many models of substellar atmospheres like the Sonora Bobcat (Marley et al. 2021), Sonora Cholla (Karalidi et al. 2021), ATMO2020 models (Phillips et al. 2020), and those generated in Mukherjee et al. (2022), are restricted to simulating cloudless objects and do not extend into the regime of Y dwarfs and temperate giant planets with effective temperatures below 400 K, where water clouds could form. Other models like those in Saumon & Marley (2008), BT-Settl (Allard et al. 2011), and Sonora Diamondback (Morley et al. 2024) include refractory clouds (e.g., iron and silicates) that are limited to objects warmer than the coldest Y dwarfs like WISE 0855. Sonora Elf Owl (Mukherjee et al. 2024) provides models for cloudless objects down to 275 K. The most up-to-date grid of models that include water clouds are published in Lacy & Burrows (2023); they produced a new grid of models for Y dwarfs that include both water clouds and disequilibrium chemistry using the *coolTLUSTY* framework (Hubeny & Lanz 1995; Sudarsky et al. 2005; Burrows et al. 2008). However, this grid of models and other studies that have treated water clouds (e.g., Morley et al. 2014a, 2014b, 2018) use cloud models that do not consider the microphysics of cloud formation.

### 1.3. Parameterized and Microphysical Models of Clouds

One method for generating cloud models is through the use of `EddySed` (Ackerman & Marley 2001), a parameterized cloud model that calculates cloud mass and particle size distributions under the assumption of horizontal homogeneity. `EddySed` has been widely utilized in various studies (e.g., Marley et al. 2010; Morley et al. 2015; Skemer et al. 2016; Rajan et al. 2017), as well as in suites of models such as those developed by Saumon & Marley (2008) and Sonora Diamondback models (Morley et al. 2024). In `EddySed`, the sedimentation efficiency parameter, $f_{sed}$, governs the vertical extent of the cloud by regulating the mixing strength of cloud particles in the atmosphere. A larger $f_{sed}$ generates optically thinner, more vertically constrained clouds with large particles, while a smaller $f_{sed}$ generates optically thicker, more vertically extended clouds with small particles.

Another more complex model, CARMA (1D Community Aerosol and Radiation Model for Atmospheres; Turco et al. 1979; Toon et al. 1988; Jacobson et al. 1994; Ackerman et al. 1995; Gao et al. 2018), provides a microphysical approach to cloud formation, considering individual processes like nucleation, coagulation, evaporation, and condensation. While complex cloud models like CARMA offer the highest level of accuracy for atmospheric simulations, their computational cost can be prohibitive. Producing a cloud profile using

CARMA may require up to 12 hr, whereas the `EddySed` cloud model (Ackerman & Marley 2001) generates cloud profiles in seconds, making it a more computationally efficient alternative.

Mang et al. (2022) investigated the impact of microphysical processes on water cloud formation in substellar atmospheres. By comparing microphysically calculated cloud profiles with parameterized cloud models for a range of Y dwarfs and temperate giant planets, Mang et al. (2022) found that the largest differences in model spectra are in the peak flux at 4–5 $\mu$m (M band). A single $f_{sed}$ value could not fully replicate the optical depth profile of the microphysical cloud model. Notably, the CARMA optical depth profile transitions from high to low $f_{sed}$ values from the top of the cloud to the cloud base. For temperate giant planets with infalling meteoritic dust and photochemical haze acting as cloud condensation nuclei, the CARMA model generated much more vertically extended cloud profiles that did not match any parameterized cloud model. They also found a large difference in the cloud base treatment where `EddySed` immediately stops cloud formation below the cloud base while CARMA has a gradual falloff.

The CARMA models in Mang et al. (2022), which were generated using cloud-free pressure–temperature (P-T) profiles, ignore the radiative feedback of the cloud on the P-T profile. Without the inclusion of radiative feedback, a spectrum generated using a cloud-free thermal structure with the additional opacity from the post-processed CARMA cloud would significantly lower the effective temperature. Including radiative cloud feedback increases the local temperature of the atmospheric layers below the cloud deck in the thermal structure, changing the cloud morphology and optical depth profile. To improve the accuracy of the model, one would have to iteratively solve for the thermal structure of the atmosphere with the cloud.

Latent heat release from the condensation of water may also affect the temperature structure, as implemented in Tang et al. (2021). When water clouds form, they release latent heat due to the phase change from gas to solid; when they sediment in the atmosphere and evaporate, they absorb heat from the surrounding material. Because condensation and evaporation occur at different rates in different layers, this could plausibly change the P-T profiles in radiative–convective equilibrium. This phenomenon has previously been explored in the context of brown dwarfs in convective regions, as detailed in Tang et al. (2021), where it changes the adiabat from a "dry" to a "moist" adiabat. In their study, it was observed that water latent heat exerts a significant influence on the thermal structure of substellar objects with effective temperatures below 350 K and supersolar metallicities. In all models generated in Tang et al. (2021), latent heat release was only accounted for in the convective regions but not the radiative regions.

### 1.4. This Work

In this study, our objective is to integrate findings from microphysical calculations into `EddySed` to develop self-consistent atmospheric models featuring water-ice clouds, while preserving crucial microphysical characteristics. We specifically address the key factors highlighted by Mang et al. (2022) that drive differences in water cloud structure and simulated spectral features, focusing on the sedimentation efficiency parameter ($f_{sed}$), cloud base treatment, and particle sizes. We explore the impact of these parameters on self-consistent atmospheric models by using CARMA models to inform modifications to `EddySed`. Additionally, since this version of the self-consistent modeling framework uses a time-stepping scheme and incorporates water latent heat release (Tang et al. 2021), we integrate microphysical particle mass growth and loss rates derived from CARMA to provide a physical constraint on cloud evolution. Section 2 details the atmospheric models employed. Section 3 presents the differences in cloud structure and their observational implications. We then discuss the broader implications of our work and suggest future improvements to modeling water clouds in substellar objects in Section 4. We conclude with a summary of our key insights in Section 5.

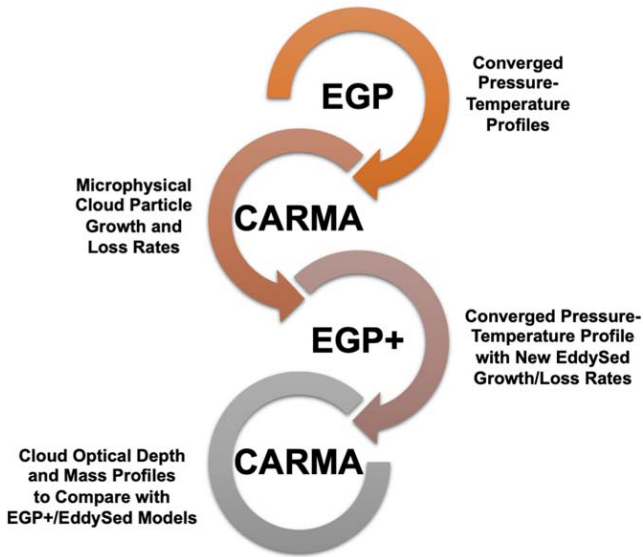

**Figure 1.** Illustration depicting the workflow and interconnections between the `EGP`, `EGP+`, and CARMA models used in this work.

## 2. Models

We use three different modeling frameworks in this study; in this section, we describe these models and how they interact with each other (Figure 1). The foundation of this work is built on the modeling framework, `EGP` (Extrasolar Giant Planet; Section 2.1) used for Solar System objects (McKay et al. 1989; Marley & McKay 1999), exoplanets (e.g., Fortney et al. 2005, 2007, 2008; Marley et al. 2012; Morley et al. 2017; Fortney et al. 2020), and brown dwarfs (e.g., Marley et al. 1996; Morley et al. 2014b; Karalidi et al. 2021; Marley et al. 2021; Zhang et al. 2021). First, we generate the entire grid of atmospheric profiles with `EGP` to be used as the starting profile in CARMA (Section 2.2). From these initial CARMA models, we calculate the microphysical rates of particle mass growth and loss and use them as inputs in `EGP+` (Section 2.3) to improve the physical constraints on the cloud evolution given `EGP+`'s time dependence. We use this updated version of `EGP+` to generate new, self-consistent models for our set of brown dwarf atmospheric profiles with water-ice clouds present.

For a direct comparison of the cloud morphology, we use the self-consistent pressure–temperature profiles from the newly upgraded `EGP+` as the thermal structure profile in CARMA to generate a second set of cloudy microphysical models. Since CARMA is not self-consistent, we use the second suite of CARMA cloud profiles to inform a new prescription for the cloud base and include the variable $f_{sed}$ in `EddySed`, aiming to reduce discrepancies in the observables when `EddySed` is used in `EGP+`. Finally, we use PICASO (Section 2.5) to simulate thermal emission spectra for all our models to evaluate the observable impact of these modifications to `EddySed`.

### 2.1. EGP

The fundamental model used in this work is the one-dimensional thermal structure model, `EGP`, which has been used in numerous works on brown dwarfs (e.g., Saumon & Marley 2008; Morley et al. 2012; Robinson & Marley 2014; Morley et al. 2018; Karalidi et al. 2021; Marley et al. 2021; Tang et al. 2021). The clouds in this model follow the `EddySed` framework described in Ackerman & Marley (2001). The model condenses all excess water vapor beyond the saturation vapor pressure above the atmospheric layer where the P-T profile crosses the condensation curve. Using the $f_{sed}$ value to parameterize the sedimentation of the cloud condensates, `EddySed` balances the downward transport of the condensate with the upward turbulent mixing of condensate and vapor,

$$-K_{zz}\frac{\partial q_t}{\partial z} = f_{sed} w_* q_c, \qquad (1)$$

where $K_{zz}$ is the eddy diffusion coefficient, $q_t$ is the condensate and vapor mixing ratio, $q_c$ is the condensate mixing ratio, and $w_*$ is the convective velocity. A single `EGP` model for a cloudy substellar atmosphere will converge within 20 to 40 minutes. During this time, the `EddySed` cloud model is called a few hundred times and almost instantaneously solves for the cloud profile.

### 2.2. CARMA

To incorporate the microphysical treatment of cloud formation in `EddySed`, we generate models using the 1D cloud microphysics model CARMA (Community Aerosol and Radiation Model for Atmospheres; Turco et al. 1979; Toon et al. 1988; Jacobson et al. 1994; Ackerman et al. 1995; Gao et al. 2018). CARMA calculates the microphysics of cloud formation, particle sizes, and vertical distributions. CARMA computes the nucleation, coagulation, condensation, and evaporation rates for each particle size bin across the atmospheric layers, and also considers the vertical mixing and sedimentation of cloud particles.

Using the self-consistent, cloudy P-T profiles from `EGP` and `EGP+`, detailed in the following section, we use CARMA to generate water-ice clouds. The profiles from `EGP` are used for the first set of CARMA models to calculate particle growth and loss rates, and `EGP+` P-T profiles are used for the final comparison of differences in treatment in cloud formation. The water-ice clouds in CARMA undergo heterogeneous nucleation onto KCl particles where these KCl particles serve as cloud condensation nuclei (CCN) and are allowed to form a homogeneous KCl cloud lower in the atmosphere from the available KCl vapor. The KCl particles are mixed upward from this cloud deck (Lodders 1999; Morley et al. 2012; Mang et al. 2022). In some cases aimed at investigating differences in cloud mass profiles, we introduce infalling meteoritic dust that is allowed to grow homogeneously through coagulation as an additional source of CCN. For the water abundance in the CARMA model, we assume a water vapor mixing ratio of $\sim 8 \times 10^{-4}$, the expected water abundance at solar metallicity and C/O ratio, that is mixed upward when the model begins to run.

Each CARMA model requires approximately 8 to 12 hr to compute a single static water-ice cloud model. Given the need for iterative models to calculate the radiative–convective profile self-consistently with clouds, CARMA is too computationally expensive to include in, e.g., `EGP`/`EGP+`. In contrast, the `EddySed` cloud model, when employed within the `EGP`/`EGP+` frameworks, allows for nearly instantaneous cloud calculations.

### 2.3. EGP+

Robinson & Marley ([2014](#)) introduced a time-stepping scheme to `EGP` and Mayorga et al. ([2021](#)) further developed it to include solar fluxes, resulting in `EGP+`. In every layer of the atmosphere, the temperature is perturbed at each time step based on the heating rate, $Q_i$, and size of the time step, $\Delta t$, expressed as

$$T_i(t + \Delta t) = T_i(t) + Q_i \Delta t, \qquad (2)$$

where $Q_i$ is based on the net energy flux, $F_{net}$, expressed as

$$Q = \frac{\partial T}{\partial t} = \frac{g}{c_p} \frac{\partial F_{net}}{\partial p}. \qquad (3)$$

The time-stepping thermal structure framework generates well-converged P-T profiles in the upper atmosphere and allows for the consideration of radiative cloud feedback. Moreover, Tang et al. ([2021](#)) integrated a moist adiabat and latent heat release into `EGP+`. While they included models for partly cloudy atmospheres, in this work we will be studying the effects of latent heat release in the radiative zone on fully cloudy atmospheres. Given the time-dependent nature of cloud formation dynamics, we can use the temporal capability of this thermal structure model to see how clouds form and dissipate from radiative feedback, in particular in light of future time-series observations of these objects.

### 2.4. EddySed Improvements

While `EddySed` tracks how condensable materials are lofted and how they fall through the atmosphere with time, it does not prescribe a timescale for particle growth. In the current `EGP+` model, water clouds instantaneously form and evaporate at each time step as the thermal structure profile changes, which is not physically realistic. In CARMA, nucleation and other microphysical processes like the growth of particles due to condensation and coagulation, or loss due to evaporation, do not occur instantaneously, and their rates can be calculated as detailed in Gao et al. ([2018](#)).

The particle distribution in CARMA is discretized into bins with associated particle size ($r_p$ [$\mu$ m]) and particle mass ($m_p$ [g]). The rate of change in each bin of the grid is controlled by nucleation ($\frac{dn}{dt}_{nuc}$), condensation ($\frac{dn}{dt}_{cond}$), coagulation ($\frac{dn}{dt}_{coag}$), and evaporation ($\frac{dn}{dt}_{evap}$), all in particles cm$^{-3}$ s$^{-1}$. For each of these processes, the production is based on growth from a smaller radius bin and shrinking particles from a larger radius bin, and vice versa for loss.

The net rate of each process ($x$) in each bin ($n$), at each altitude ($z$), is the following:

$$\frac{dn}{dt}_{x,net}(n, z) = \frac{dn}{dt}_{x,prod}(n, z) - \frac{dn}{dt}_{x,loss}(n, z). \qquad (4)$$

To determine the maximum allowable particle mass growth based on the CARMA microphysics at each layer, we use the nucleation and condensation rates. The growth rate at each layer will therefore be

$$\frac{dm}{dt}_{growth}(z)$$
$$= \sum_{n=1}^{nbin} \left( \frac{dn}{dt}_{nuc,net}\left(n, z\right) + \frac{dn}{dt}_{cond,net}(n, z) \right) * m_p(n), \qquad (5)$$

where the growth rate at each layer is in grams cm$^{-3}$ s$^{-1}$.

For the particle mass-loss rate, we only have evaporation where the loss rate in each layer is

$$\frac{dm}{dt}_{loss}(z) = \sum_{n=1}^{nbin} \frac{dn}{dt}_{evap,net}(n, z) * m_p(n, z). \qquad (6)$$

The rate of change in condensate mass mixing ratio ($\frac{dq_c}{dt}$ [g/g/s]) is defined as

$$\frac{dq_c}{dt}(z) = \frac{q_c - q_{c,0}}{dt}, \qquad (7)$$

where $q_{c,0}$ is the condensate mass mixing ratio value from the previous time step.

We can then calculate the threshold for the change in condensate mass mixing ratio as informed by the CARMA rates to be used in `EGP+` with the following:

$$\frac{dq_c}{dt}_{growth}(z) = \frac{dm}{dt}_{growth} * \frac{1}{\rho_{atm}(z)}, \qquad (8)$$

where $\rho_{atm}$ is the mass density of a layer in the atmosphere in grams cm$^{-3}$, and

$$\frac{dq_c}{dt}_{loss}(z) = \frac{dm}{dt}_{loss} * \frac{1}{\rho_{atm}(z)}. \qquad (9)$$

In each iteration, we compare the `EddySed` calculated value of $\frac{dq_c}{dt}$ to our CARMA-prescribed growth and loss rates. If $\frac{dq_c}{dt}(z)$ is positive, it is compared to $\frac{dq_c}{dt}_{growth}$. If it is negative, then it is compared to $\frac{dq_c}{dt}_{loss}$. If $\frac{dq_c}{dt}(z)$ exceeds either of these limits, $\frac{dq_c}{dt}(z)$ is set to the respective limiting CARMA-derived value. Consequently, $q_c$ is adjusted as follows:

$$q_{c,new} = \frac{dq_c}{dt}_{growth}(z) * dt + q_{c,0}. \qquad (10)$$

We use these rates to prescribe thresholds in `EddySed` to generate microphysically motivated cloud mass growth and loss rates in the time-stepping scheme. We then monitor the temporal cloud optical profile evolution. We also compare the CARMA water-ice cloud particle sizes, size distributions, and optical properties, with those from the `EddySed` model.

In contrast to prior models that adhered to a constant $f_{sed}$ value per the Ackerman & Marley ([2001](#)) cloud treatment (Morley et al. [2014a](#), [2018](#); Mang et al. [2022](#)), Rooney et al. ([2022](#)) introduced a new formulation for $f_{sed}$ that allows it to vary across the atmospheric profile. Using a variable $f_{sed}$ parameter becomes imperative for achieving better alignment of cloud profiles with microphysical cloud models (Mang et al. 2022; Rooney et al. 2022). We incorporate the Rooney et al. (2022) approach into the EddySed framework within EGP+ for this work, resulting in better agreement between the optical depth profiles computed by EddySed and CARMA.

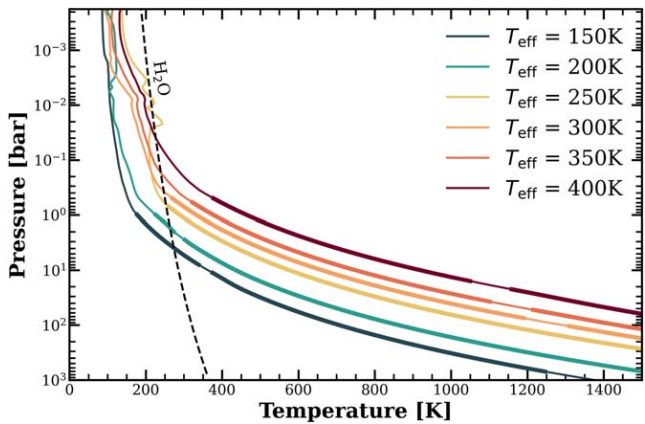

**Figure 2.** Pressure–temperature profiles for a range of effective temperatures with a surface gravity of $\log(g) = 4.5$ (cgs). The condensation curve of water ice is shown in the black dashed line. The thicker regions of the profile indicate where the atmosphere is convective.

### 2.5. PICASO

We employ PICASO to generate moderate-resolution thermal emission spectra for all of our models (Batalha et al. 2019; Batalha & Rooney 2020; Batalha et al. 2021). We use the P-T profiles and abundances from EGP+, and post-process the spectra at higher resolution. The spectra cover a range from 0.6 to 15 μm at a resolving power of 500. These spectra are computed using the opacity database from Freedman et al. (2008), with significant updates as described in Freedman et al. (2014). Furthermore, we incorporate the latest revised opacity table outlined in Marley et al. (2021), which includes enhancements to $H_2O$, $CH_4$, $FeH$, and the alkali metals.

### 2.6. Model Parameters

We compute the P-T profiles for brown dwarfs assuming solar metallicity and C/O ratio with chemical equilibrium abundances based on Lodders & Fegley (2002, 2006) and Visscher et al. (2006, 2010). The vertical mixing strength in the atmosphere is parameterized by the eddy diffusion coefficient ($K_{zz}$). This is computed using mixing length theory with a minimum value of $10^5\,\mathrm{cm^2\,s^{-1}}$ (Ackerman & Marley 2001; Morley et al. 2014b). The impacts of these assumptions are discussed in Section 4.

We use converged EGP+ atmospheric P-T profiles to generate CARMA models to facilitate a direct assessment of disparities in the two cloud formation models (Figure 2). We generate a grid of models with $T_{\mathrm{eff}} = [150, 200, 250, 300, 350, 400\,\mathrm{K}]$ and $\log(g) = [3.5, 4.0, 4.5, 5.0]$, in cgs units. We also generate a 175 K, $\log(g) = 4.5$ model aimed at investigating large differences found between 150 K and 200 K models. In all cases, we use a moist adiabat with water latent heat release enabled in both the convective and radiative regions. For both CARMA and EddySed, we use Mie theory to compute cloud optical properties, including the optical depth, single scattering albedo, and asymmetry parameter.

For the CARMA models, we calculate water-ice clouds formed through homogeneous and heterogeneous nucleation, where KCl serves as the condensation nuclei, following the cloud treatment outlined in Mang et al. (2022). To explore the potential limitations imposed by condensation nuclei in Section 3.5, we also generate models with water-ice clouds nucleating on meteoritic dust particles falling from the upper atmosphere. We use downward fluxes of $10^{-19}$ and $10^{-17}\,\mathrm{g\,cm^{-2}\,s^{-1}}$, as Mang et al. (2022) observed negligible variations in the optical depth profiles with higher fluxes. Models presented in this work are publicly available on Zenodo (Mang et al. 2024).

## 3. Results

We first compare the cloud profiles generated here to those from Mang et al. (2022). Figure 3 shows the difference between the CARMA cloud profiles generated with the cloud-free EGP thermal structure utilized in Mang et al. (2022) and the corresponding cloudy EGP+ thermal structure developed in this work. In the cloudy EGP+ model, we integrated the microphysical cloud growth and loss rates derived from CARMA (Section 3.4). Due to the presence of clouds in the atmosphere of the EGP+ thermal structure, there is a slight elevation in temperature within the layers where cloud condensation occurs compared to the cloud-free EGP profile. Consequently, upon post-processing a water-ice cloud using CARMA, we observe an order of magnitude decrease in the peak optical depth of the cloud.

Next, we discuss five key results from our joint microphysics and thermal structure models of Y dwarfs. First, we investigate the presence of supersaturation in the CARMA models in Section 3.1. Then, we evaluate the optical depth profile using a variable $f_{\mathrm{sed}}$ in comparison to the CARMA cloud model in Section 3.2. In Section 3.3, we introduce a new cloud base prescription by assessing the optical depth falloff rate of the CARMA cloud model and its effect on the brightness temperature profiles. Section 3.4 presents an examination of the particle size distribution, the microphysical rates for our grid of models, and their effects on the thermal emission spectra. In Section 3.5, we examine the cloud condensation nuclei and their role in limiting cloud formation. Finally, we explore the impact of latent heat release on water cloud formation in Section 3.6.

### 3.1. Supersaturation

The current EddySed framework condenses all water vapor in excess of the saturation vapor pressure. Supersaturation serves as a critical condition that, when reached in the atmosphere, enables the condensation of water vapor onto CCNs, facilitating the creation of cloud droplets. In the Earth's troposphere, supersaturation up to 160% relative humidity can occur with respect to ice (Jensen et al. 2001).

In the CARMA models, some water vapor in excess of the saturation vapor pressure can remain in the gas phase, in contrast to the current prescription in EddySed where 100% of the excess water vapor condenses (Figure 4). For the coldest two cases with effective temperatures of 150 and 200 K, the CARMA models condense 98.5% to 99.9% of the water vapor, resulting in efficient water-ice cloud formation and removal of water vapor. As the models become warmer, the thermal structure profile sits closer to the water condensation curve; at

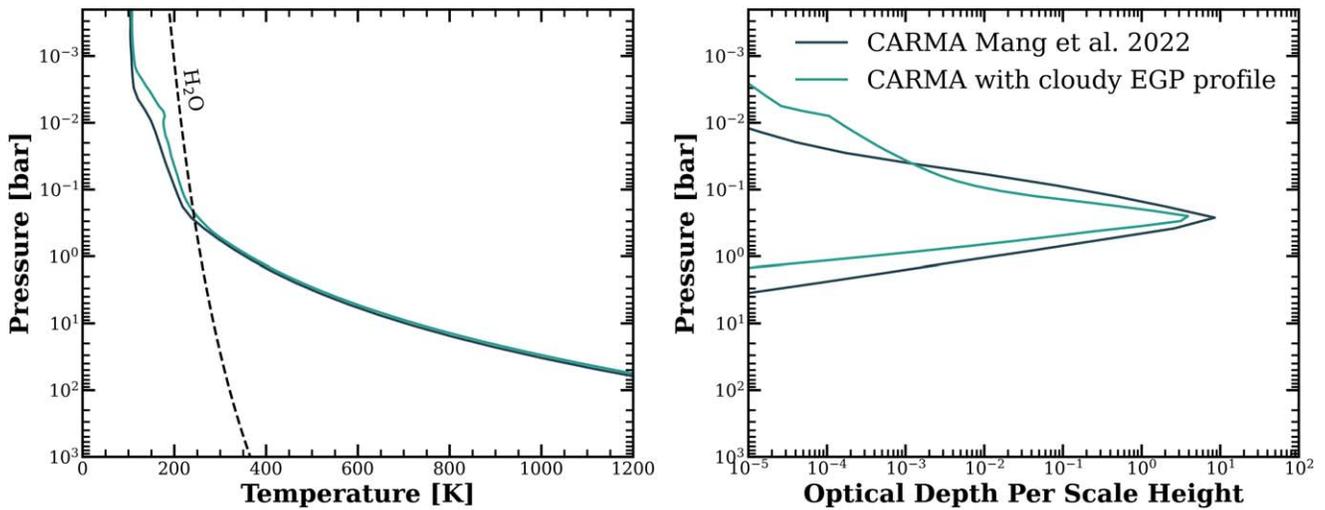

**Figure 3.** Left: The cloud-free `EGP` P-T profile used in Mang et al. (2022) in comparison to the cloudy ($f_{\rm sed} = 6$) `EGP`+ P-T profile generated in this work for a Y dwarf with $T_{\rm eff} = 350$ K, $\log(g) = 4.5$ (left). Right: The optical depth profile of the CARMA cloud using these thermal structures.

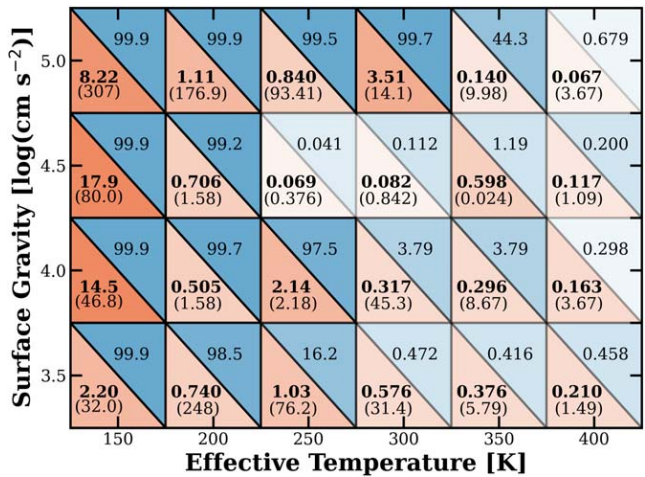

**Figure 4.** Summary of the optical depth at 4.074 $\mu$m at the cloud base of the CARMA cloud model (in bold) and `EddySed` (in parentheses) in the red, and the percentage of the water that is condensed in the CARMA cloud model in blue, for our grid of models. Darker shades represent higher values, while lighter shading indicates lower values.

250 K, we see a transition to less efficient cloud formation. For models from 350 to 400 K, only 0.2%–44.3% of the excess water is actually condensed, making the atmosphere supersaturated in water vapor.

In all cases, the CARMA clouds are less optically thick than the `EddySed` cloud models with $f_{\rm sed} = 6$. This is due to two factors: in part, it is due to the higher supersaturation of the CARMA clouds (and therefore a lower mass of condensed material). It is also largely due to the difference in particle size distributions, as the `EddySed` cloud particles are generally much smaller than the CARMA particles, which is discussed further in Section 3.4. Although the atmospheres in our grid are supersaturated, the homogeneously nucleated water clouds in this work exhibit optical depths orders of magnitude smaller than those of heterogeneously nucleated water clouds, consistent with the findings of Mang et al. (2022). As a result, the homogeneously nucleated water clouds closely resemble cloud-free models.

The 250 K, $\log(g) = 4.5$ case (and other less dramatic nonlinearities in the results across our grid) showcases the difficulty of the convergence of fully cloudy models as this P-T profile fluctuates around the condensation curve (Figure 2), making it more unstable for cloud condensation. A slightly different P-T profile may have a thicker cloud at this effective temperature.

We can adjust the water vapor that condenses in the `EddySed` model to better match our results from the CARMA microphysical models. We use two approaches within `EddySed`. The first method involves running `EddySed` models with different levels of supersaturation after condensation. The second approach entails adjusting the condensate mass mixing ratio in the `EddySed` cloud model.

The supersaturation after condensation is parameterized by $S_{\rm cloud}$ (Marley et al. 1999; Ackerman & Marley 2001):

$$S_{\rm cloud} = \frac{q_v(z - \Delta z) - q_c(z)}{q_s(z)} - 1, \tag{11}$$

where $q_v$ is the vapor mass mixing ratio, $q_c$ is the condensate mass mixing ratio, and $q_s$ is the saturation vapor mass mixing ratio. For example, an $S_{\rm cloud}$ value of 0.1 represents a 10% increase in the saturation vapor pressure, increasing the barrier for additional cloud formation beyond the cloud that has already condensed. Note that $S_{\rm cloud}$ is different from $S$, supersaturation before condensation, as discussed in Ackerman & Marley (2001).

We examined three different $S_{\rm cloud}$ values. We selected 0.1 and 1 based on the extreme cases presented in Ackerman & Marley (2001), while also introducing 10 as an additional case derived from the CARMA models. Despite a reduction in the total column mass, the column mass profile using $S_{\rm cloud} = 0.1$ and 1 fails to replicate the CARMA profile (Figure 5). Notably, the `EddySed` cloud column mass for the $S_{\rm cloud} = 10$ case exhibits the closest alignment with the CARMA model. The elevated cloud base observed for the $S_{\rm cloud} = 10$ case, in comparison to the other models, is due to the water condensation curve shifting to the left in the P-T profile, as increasing $S_{\rm cloud}$ increases the threshold at which the vapor can condense. Consequently, this places the intersection of the condensation curve and the thermal structure of the object (e.g., cloud base) farther up in the atmosphere.

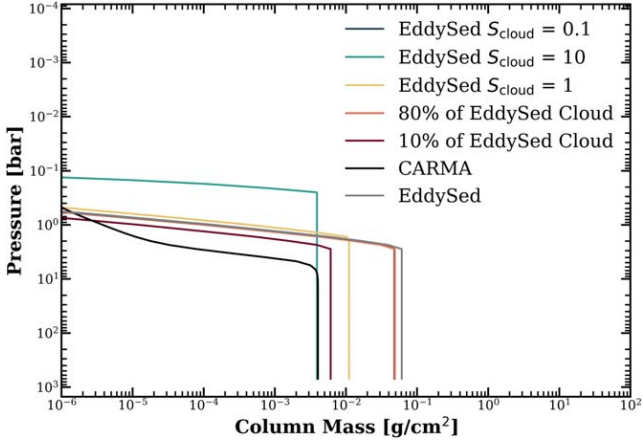

**Figure 5.** Water cloud column mass of a CARMA model and EddySed models with various $S_{cloud}$ values and different fractions of the cloud mass mixing ratio for a Y dwarf with $T_{eff} = 175$ K, $\log(g) = 4.5$ and $f_{sed} = 6$.

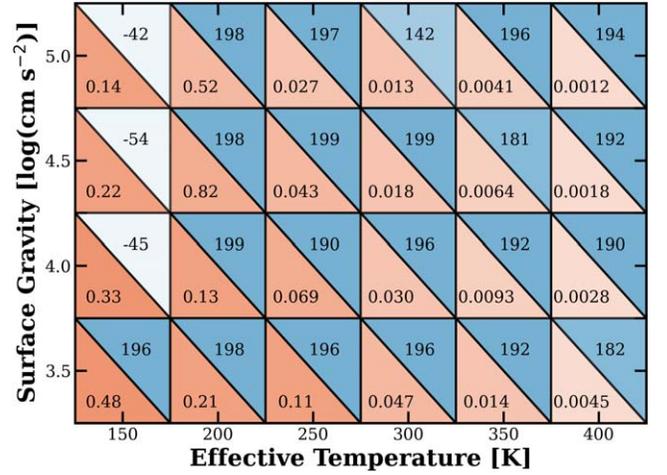

**Figure 6.** Summary of the maximum (EddySed) cloud column mass in units of g cm$^{-2}$ in red and the percentage difference with the CARMA cloud column mass (Equation (12)) in blue, for our grid of models with $f_{sed} = 6$. Darker shades represent higher values, while lighter shading indicates lower values.

Alternatively, we modify the EddySed water cloud by multiplying the condensate mass mixing ratio by 0.1 and 0.8 to test 10% and 80% cloud fractions, essentially removing 90% and 20% of the EddySed cloud. This maintains the position of the condensation curve and a closer match to the CARMA model's column mass is achieved, as illustrated in Figure 5. For this specific case with $T_{eff} = 175$ K, reducing the condensate mass mixing ratio aligns the cloud base at the same level as the CARMA model. However, for objects with temperatures higher than this threshold, adjusting the fraction of condensate mass mixing ratio also influences where the cloud base is. As previously discussed, as these objects get warmer, their thermal structures approach the water condensation curve, making the P-T profile highly sensitive to any changes to the water clouds. Therefore, taking any fraction of the original EddySed cloud changes the thermal structure of the object and causes the cloud base to be farther up in the atmosphere, similar to the effects of using $S_{cloud} = 10$.

Figure 6 provides an overview of the cloud column mass difference between CARMA and EddySed models, highlighted in blue, across the entire grid, where the percentage difference, %$\Delta$, is calculated as follows:

$$\%\Delta = \frac{m_{col, EddySed} - m_{col, CARMA}}{m_{col, EddySed}} \times 100. \quad (12)$$

A noticeable transition occurs from the 150 K models to the 200 K models, where a reduced amount of condensation of water clouds in the latter is evident. This is further illustrated by the significant difference in the optical depth of the corresponding three 150 K models with $\log(g) = 4.0$, 4.5, and 5.0, as shown in Figure 4. These three cases have more vertically extended CARMA cloud profiles with extremely efficient cloud condensation compared to the other models in the grid, where the supersaturation is just high enough to enable substantial water cloud formation.

We evaluate the effects of variations in modeled cloud properties on spectra in Figure 7, which displays the simulated spectra for all cases with reduced column mass in Figure 5. The adjustment of cloud column mass to align with the CARMA profile does not necessarily result in a closer match to the spectra. This may be due to the large difference in particle size distributions where the cloud mass between CARMA and

EddySed models can be more closely aligned but the particle size distribution drives the optical properties dominating the effects on the spectra. We further discuss particle size differences in Section 3.4. The models with $S_{cloud}$ values of 0.1 and 1 exhibit negligible differences compared to the original EddySed model. The $S_{cloud} = 10$ model is essentially cloud-free but is brighter in small windows due to increased flux from small isothermal regions in the thermal structure that dictates the amount of flux in these spectral regions. Notably, the CARMA model's spectrum falls between the spectra of $S_{cloud} = 1$ and 10. The models with a fraction of the EddySed cloud either become too transparent or remain too optically thick, as seen in the 3 to 6 μm region.

Although both approaches effectively reduce the cloud column mass in EddySed to align more closely with the CARMA water cloud column mass profile, the simulated thermal emission spectra suggest a more substantial impact from the different optical properties of the water cloud. Changing the water cloud column mass also induces variations in the thermal structure across the majority of the grid. This directs our attention toward achieving a better match between EddySed and the CARMA models by exploring the impacts of differences in optical depth profiles and particle size distributions, as detailed in Sections 3.2 and 3.4.

### 3.2. Variable $f_{sed}$

To compare EddySed to our full CARMA grid, we calculate the best-fit variable $f_{sed}$ prescription to the cloud optical depth profile. The variable $f_{sed}$ is calculated as follows:

$$f_{sed}(z) = \alpha \exp\left(\frac{z - z_T}{6\beta H_0}\right) + \epsilon, \quad (13)$$

where $\alpha$ is a constant of proportionality, z is the altitude, $z_T$ is the altitude of the top layer of the atmosphere, $\beta$ is the scaling parameter, $H_0$ is the scale height, and $\epsilon$ is the minimum allowable $f_{sed}$ value, which we set to 0.01 following the minimum value found for super-Earths in Morley et al. (2015). When utilizing a variable $f_{sed}$ parameter with a suite of $\alpha$ and $\beta$ values, $\alpha$ is the maximum $f_{sed}$ value we derive from the constant $f_{sed}$ models that best fit the optical depth profile of the CARMA cloud, and $\beta$ determines the rate of change of $f_{sed}$.

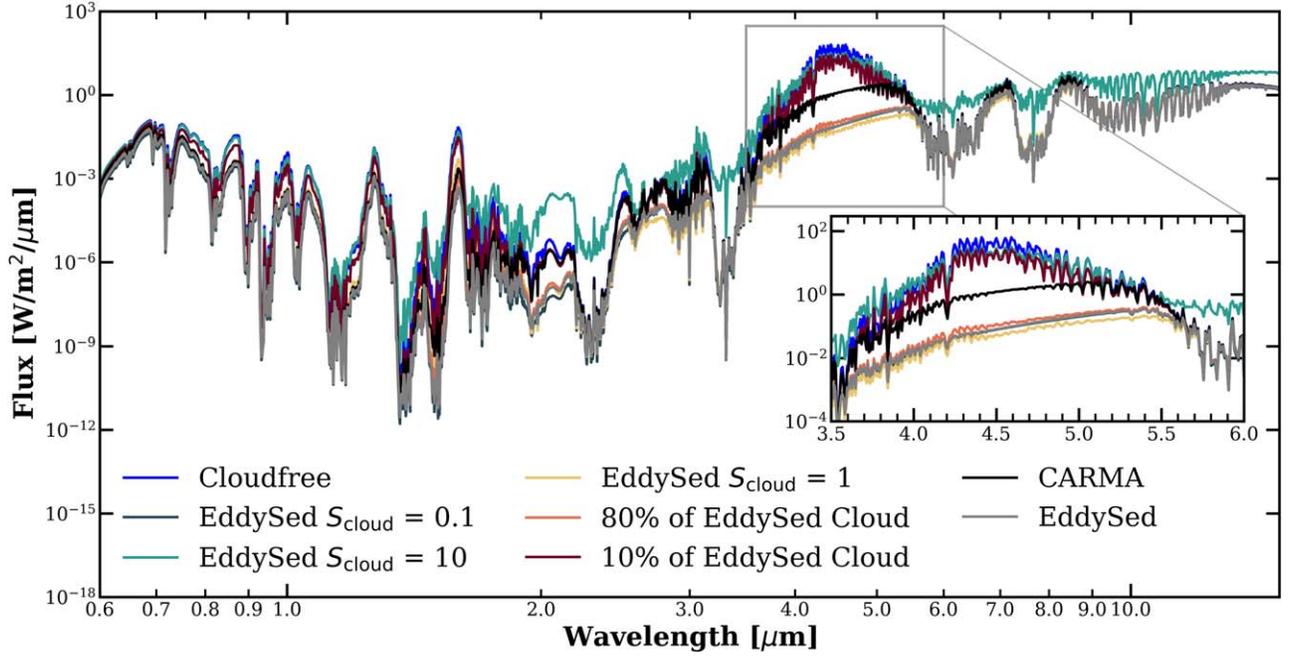

**Figure 7.** Simulated thermal emission spectra from 0.6 to 15 $\mu$m of a CARMA model, `EddySed` models with various $S_{cloud}$ values and different fractions of the cloud mass mixing ratio, and a cloud-free model for a Y dwarf with $T_{eff} = 175$ K, $\log(g) = 4.5$ and $f_{sed} = 6$. The inset highlights the region of interest between 3.5 and 6 $\mu$m.

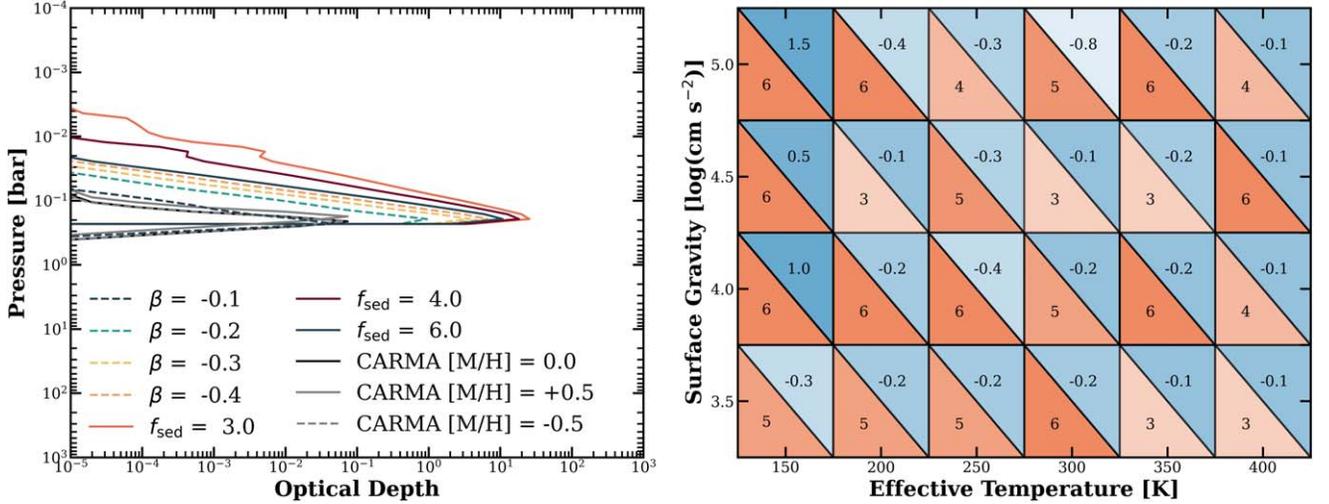

**Figure 8.** Left: Comparison of CARMA and `EddySed` water cloud optical depth profiles for a Y dwarf with $T_{eff} = 300$ K, $\log(g) = 4.5$ at 4.074 $\mu$m. Both constant (solid) and variable (dashed) $f_{sed}$ `EddySed` models are shown. The starting $f_{sed}$ value ($\alpha$) for all the variable models with different $\beta$ values is 3. The best-fitting model with $\beta = -0.1$ includes the gradual cloud base falloff detailed in Section 3.3. Right: Summary of the recommended $f_{sed}$ values (red) and $\beta$ values (blue) for our grid of models. Darker shades represent higher values while lighter shading indicates lower values.

The use of the variable $f_{sed}$ parameter significantly improves the fit of the optical depth profiles of the `EddySed` models to those of the CARMA models (Figure 8). The improvement is particularly stark for the maximum optical depth reached near the cloud base and the vertical extent of the cloud. The negative $\beta$ value increases the $f_{sed}$ value going deeper into the atmosphere, resulting in larger particles and a more optically thin cloud. `EddySed` clouds with a constant $f_{sed}$ are typically optically thicker than the CARMA clouds, as demonstrated in Mang et al. (2022) and shown in the left panel of Figure 8. This is further supported by the condensate mass mixing ratio (mmr) and the mean particle radius profiles depicted in Figure 9. Compared to the constant $f_{sed}$ `EddySed` cloud particle radii, the best-fit variable $f_{sed}$ with $\alpha = 3$ and $\beta = -0.1$ almost perfectly matches the CARMA mean particle radius at the cloud base. This directly translates into the observables as seen in Figure 10. The fiducial `EddySed` case here with the constant $f_{sed}$ prescription is too optically thick, with a condensate mass mixing ratio that is higher by a few orders of magnitudes and a mean particle radii at the cloud base that is smaller by a couple orders of magnitude.

In cases with more vertically constrained CARMA clouds, a negative $\beta$ value yields the most accurate fit. Conversely, in scenarios where the CARMA cloud is more vertically

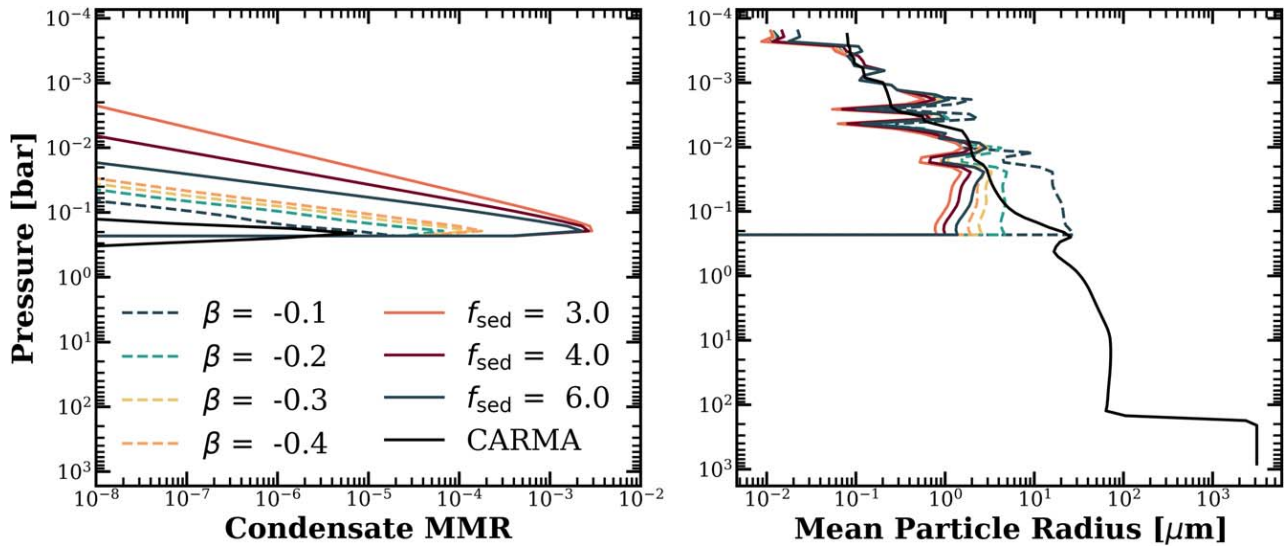

**Figure 9.** Left: Comparison of CARMA and `EddySed` water cloud condensate mass mixing ratios for a Y dwarf with $T_{eff} = 300$ K, $\log(g) = 4.5$. The constant `EddySed` models are shown (solid) along with the variable (dashed) $f_{sed}$ models that start with $\alpha = 3$. Right: The mean particle radius of the CARMA water cloud in comparison to the different `EddySed` profiles.

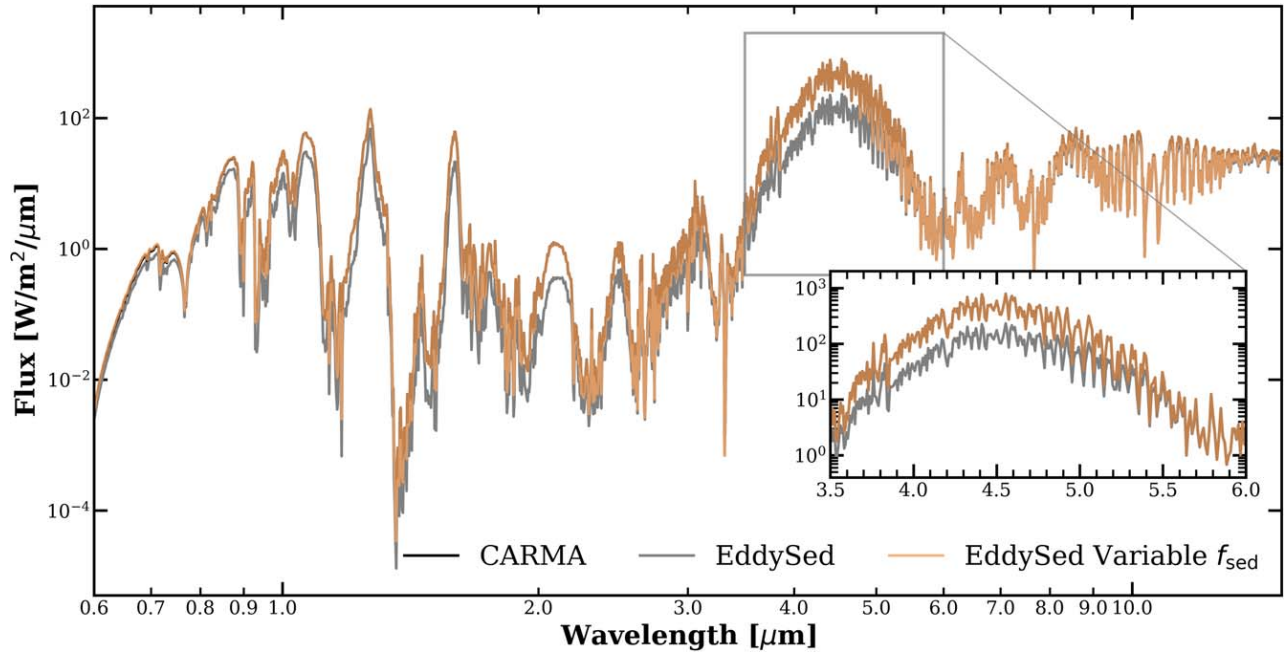

**Figure 10.** Simulated thermal emission spectra from 0.6 to 15 $\mu$m of the CARMA (black) model and `EddySed` model with a constant (gray) and variable (orange) $f_{sed}$ for a $T_{eff} = 300$ K, $\log(g) = 4.5$ Y dwarf. $f_{sed} = 3$ for the constant `EddySed` case while the variable $f_{sed}$ case corresponds to the best-fit profile in Figure 8 with $\alpha = 3$ and $\beta = -0.1$. In this case, the `EddySed` with the variable $f_{sed}$ almost directly overlaps the CARMA model. The inset highlights the region of interest between 3.5 and 6 $\mu$m.

extended, like the homogeneously nucleated water clouds discussed in Mang et al. (2022), a single variable $f_{sed}$ cloud profile cannot entirely replicate the CARMA cloud. Here, two distinct $\beta$ values are required to achieve a more accurate match, suggesting that a different parameterization of the variable $f_{sed}$ may be needed. A single negative $\beta$ value aligns the `EddySed` profile more closely to the CARMA profile near the cloud base, but it leads to insufficient optical depth in the upper cloud profile. When employing two $\beta$ values, the optimal-fit model features a positive $\beta$ value (resulting in a decrease in $f_{sed}$) in the upper half of the cloud profile, while a negative $\beta$ value

(causing an increase in $f_{sed}$) is needed in the lower half of the cloud profile.

In all instances, a constant $f_{sed}$ in `EddySed` results in a cloud that is optically thicker near the cloud base compared to CARMA. This suggests an overestimation of cloud condensation in the atmosphere as previously discussed in Section 3.1. We also find minimal differences in the optical depth profile for different metallicities. For the subsolar ([M/H] = −0.5), solar, and supersolar ([M/H] = +0.5) cases, the variable $f_{sed}$ with our recommended $\alpha$ and $\beta$ values still present the best fit in comparison to a constant $f_{sed}$.

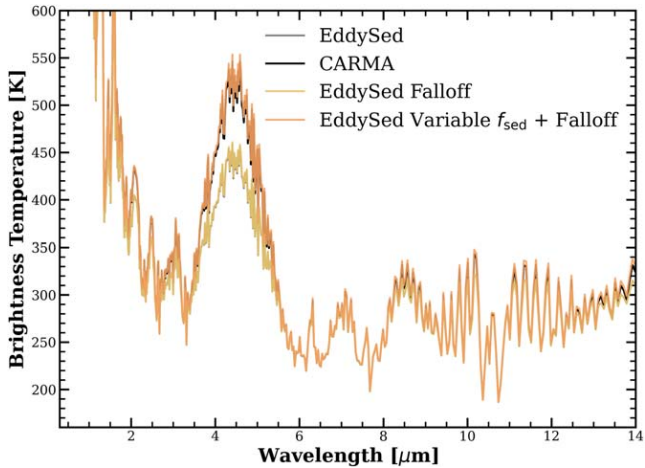

**Figure 11.** Brightness temperature as a function of wavelength for a CARMA model and corresponding `EddySed` model for a $T_{eff} = 300$ K, $\log(g) = 4.5$ Y dwarf with $f_{sed} = 3$. We apply the cloud base falloff to the original `EddySed` cloud profile (yellow) and the best-fit variable $f_{sed}$ profile with $\beta = -0.1$ (orange) as seen in Figure 8.

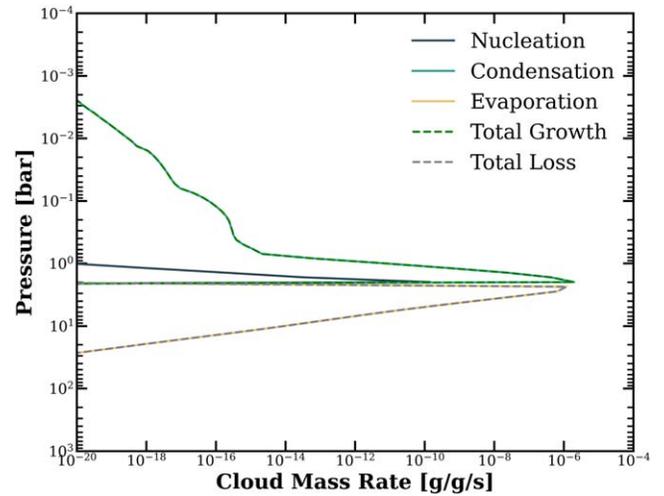

**Figure 12.** Water-ice cloud mass growth and loss rates from the CARMA model with $T_{eff} = 200$ K and $\log(g) = 4.5$. Nucleation and condensation contribute to the total growth of cloud particle mass while evaporation is the only process contributing to the total loss in particle mass. The total growth rate curve in the dashed green overlaps the condensation rate, the dominant process for particle mass growth.

To best emulate the morphology of a microphysical water cloud using `EddySed`, we recommend adopting the variable $f_{sed}$ parameter formulation introduced by Rooney et al. (2022), with $\alpha$ values falling within the range of 3–6, aligning with the best-fitting constant $f_{sed}$ values identified in Mang et al. (2022). In the right panel of Figure 8, we provide the optimal $\beta$ value near the cloud base. These serve as our recommended initial values to use for objects within this parameter space. The $\beta$ value may differ for individual models due to the large variations in the P-T profiles and thus cloud profiles from convergence issues, which are discussed in Section 4.1.

### 3.3. Cloud Base Treatment

There is a notable difference in each model's treatment of the cloud base. `EddySed` exhibits an immediate cutoff at the cloud base, as condensation ceases in the layer where the P-T profile intersects the water condensation curve. Conversely, CARMA produces a gradual decline in optical depth at the cloud base due to the finite rate of evaporation. Across our grid of CARMA models, the average falloff rate, from the cloud base down to the layer where the optical depth is $\sim 10^{-5}$, is $3.85\% \pm 2.08\%$ per layer. For a subset of models with effective temperatures of 175, 200, 250, 300, 350, and 400 K, with log $(g) = 4.5$, the falloff rates are 5.04%, 1.50%, 1.27%, 2.42%, 3.02%, and 3.14%, respectively. For a set of models with an effective temperature of 350 K, and $\log(g) = 3.5$, 4.0, and 4.5, the falloff rates are 7.36%, 4.75%, and 6.17%, respectively. No strong correlations are observed between effective temperature, surface gravity, or effective particle radius at the cloud base.

The recommended prescription for the optical depth cloud base falloff therefore is

$$\tau(P) = \tau(P_{base}) * \exp\left[-((P/P_{base})^{3.85} - 1)\right], \quad (14)$$

where $\tau$ is the optical depth and $P$ is the pressure of the atmosphere greater than $P_{base}$, the cloud base layer pressure.

By applying the average falloff rate to `EddySed`, we allow the optical depth profile below the cloud deck cutoff to more closely align with that of the CARMA model, as demonstrated in the left panel of Figure 8. Below the microphysical cloud deck, the asymmetry parameter and single scattering albedo exhibit negligible variation, so we adopt constant values based on those at the cloud base.

The impact of the cloud base falloff prescription on the observables is depicted in Figure 11. The `EddySed` cloud is more optically thick than the CARMA model and thus has a lower brightness temperature in the 3–6$\mu m$ region. With the falloff incorporated, the brightness temperature spectrum of the `EddySed` model is relatively unaffected. When the variable $f_{sed}$ and the cloud base falloff are applied, the brightness temperature profile of the `EddySed` model closely mirrors that of the original CARMA model. Since the water cloud base optical depth can differ by two orders of magnitude, depending on the $f_{sed}$ value (Figure 8), the high optical depth at the cloud base drives the observable features. Therefore, the variable $f_{sed}$ is the primary factor impacting the observable in comparison to any additional cloud contribution below the cloud base from the falloff. Initial investigations with vertically extended, very optically thin, CARMA cloud models highlighted discrepancies when the cloud base falloff is omitted. However, this has little impact on optically thick, vertically constrained clouds, such as those generated in our suite of models.

We recommend incorporating the cloud base falloff for potentially vertically extended water cloud profiles as it aligns more closely with the microphysical model predictions. For regions beneath the cloud base, we advise implementing a falloff rate of approximately 4%, starting from the cloud base optical depth value, while maintaining a constant single scattering albedo and asymmetry parameter consistent with the values at the cloud base.

### 3.4. Water-ice Cloud Particle Size and Growth Rates

Figure 12 shows the mean particle mass growth and loss rates for each microphysical process over the last half of the CARMA model run. Condensation dominates cloud particle growth above the cloud. Meanwhile, the particle mass loss is the fastest below the cloud deck, due to evaporation, since the temperature is higher than the water cloud condensation curve.

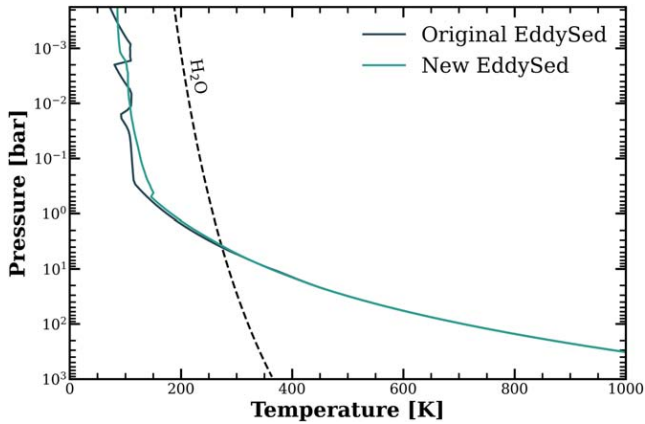

**Figure 13.** A comparison of the steady-state pressure–temperature profile of the original `EddySed` model with the new, microphysically informed `EddySed` for a Y dwarf with $T_{eff} = 175$ K and $\log(g) = 4.5$.

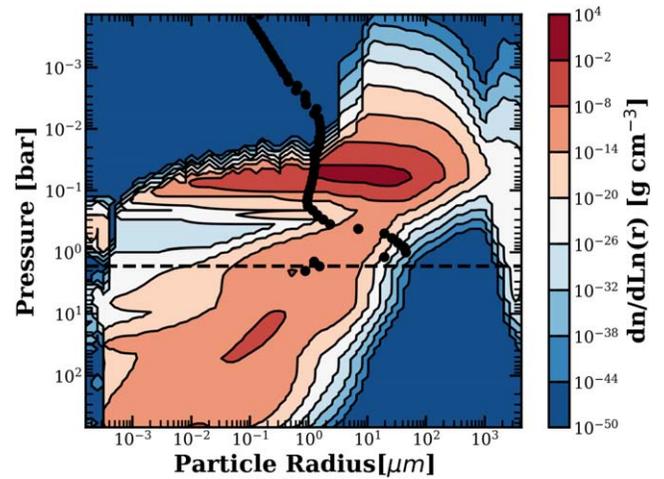

**Figure 14.** Contour plot of the CARMA water-ice cloud mass density as a function of pressure and particle radius for the $T_{eff} = 200$ K and $\log(g) = 4.5$ case. The black dashed line represents the cloud base. The `EddySed` mode cloud particle radii for a model with $f_{sed} = 6$ are shown in black circles for comparison.

The incorporation of CARMA-derived growth and loss rate in `EddySed` has resulted in increased stability in the cloud and P-T profiles as they evolve with each time step, reducing large, nonphysical cloud profile variations as we move toward a steady-state solution. The more realistic timescale of water-ice cloud evolution also improves the treatment of radiative cloud feedback and potential time-dependent cloud variability within `EGP+`. This is illustrated in Figure 13, where the P-T profile with the `EddySed` model incorporating the microphysically informed growth and loss rates in `EGP+` is much smoother and well-converged in comparison to the original `EddySed`, `EGP` profile. When modeling water clouds within a time-stepping framework, we recommend using the prescribed rates of particle growth and loss to achieve a physically informed temporal evolution of water clouds. We discuss the sensitivity of the steady-state P-T profile to the microphysical rates in the Appendix.

Comparison of the number density of cloud particles as a function of pressure and particle size between `EddySed` and CARMA (Figure 14) shows that the CARMA cloud contains larger particles in the upper atmosphere but more small particles near the cloud base. The presence of larger particles above the cloud base may be due to the lower nucleation rates in these layers of the atmosphere leading to fewer particles. The particles that are present in these layers can then grow through condensation and coagulation, a process not accounted for in `EddySed`. Condensation dominates particle growth because, after a cloud particle has formed through nucleation, it is more favorable to growth via vapor uptake rather than coagulation. Figure 14 also reveals a higher concentration of smaller particles in the lower atmosphere below the cloud, due to the large particles evaporating as they fall to deeper layers of the atmosphere.

The contrast in particle distribution between `EddySed` and CARMA is evident in both the upper and lower regions of the water-ice cloud (Figure 15). `EddySed` uses a log-normal distribution, which is unable to capture the bimodal distribution seen in CARMA. At both pressure levels shown, there are up to $10^{12}$ times more cloud particles in the `EddySed` model than in the CARMA model, depending on the choice of $f_{sed}$. This is consistently observed across the entire model grid. There are significantly more smaller particles in the constant $f_{sed}$

`EddySed` models in comparison to the CARMA model, which exhibits a flatter distribution. This results in the much greater optical depths seen in all `EddySed` cloud profiles compared to CARMA clouds, as optical depth scales with the cross-sectional area. This also aligns with the more vertically extended optical depth profile of the `EddySed` clouds. Incorporating the best-fit variable $f_{sed}$ parameter in EddySed results in a particle size distribution much closer to the modal radius observed in CARMA (Figure 15).

### 3.5. Cloud Condensation Nuclei

To further investigate the role of CCN in limiting cloud formation in CARMA models, we compare simulated thermal emission spectra with KCl particles as CCN with models that have varying amounts of infalling meteoritic dust as CCN (Figure 16). The `EddySed` spectrum exhibits considerably deeper molecular absorption and suppression of spectral features from the water-ice cloud in comparison to the various CARMA spectra. When examining the critical 3–6 $\mu$m region, even the CARMA cloud with the highest influx of dust fails to match the higher opacity of the water-ice cloud produced in the `EddySed` model. Therefore, CCN limitation contributes to the less massive microphysical clouds.

### 3.6. Latent Heat Release in the Radiative Zone

In extension to the latent heat release included in the convective regions of the atmosphere in Tang et al. (2021), we include the tracking of latent heat fluxes in the radiative regions where water clouds may be present. Because both CARMA and `EGP+` are time-stepping codes, we can use the rates of condensation and evaporation to calculate the net heat per second added to each layer by this process. We find that the contribution from the latent heat flux is minimal (Figure 17), and its impact on the P-T profile and water-ice cloud formation for models at solar metallicity is negligible. In all instances across our model grid, the total net flux conveyed in each radiative layer is approximately $10^5$ W m$^{-2}$ while the contribution from the latent heat flux spans from $10^{-2}$ to $10^{-25}$ W m$^{-2}$. This aligns with the conclusions drawn in

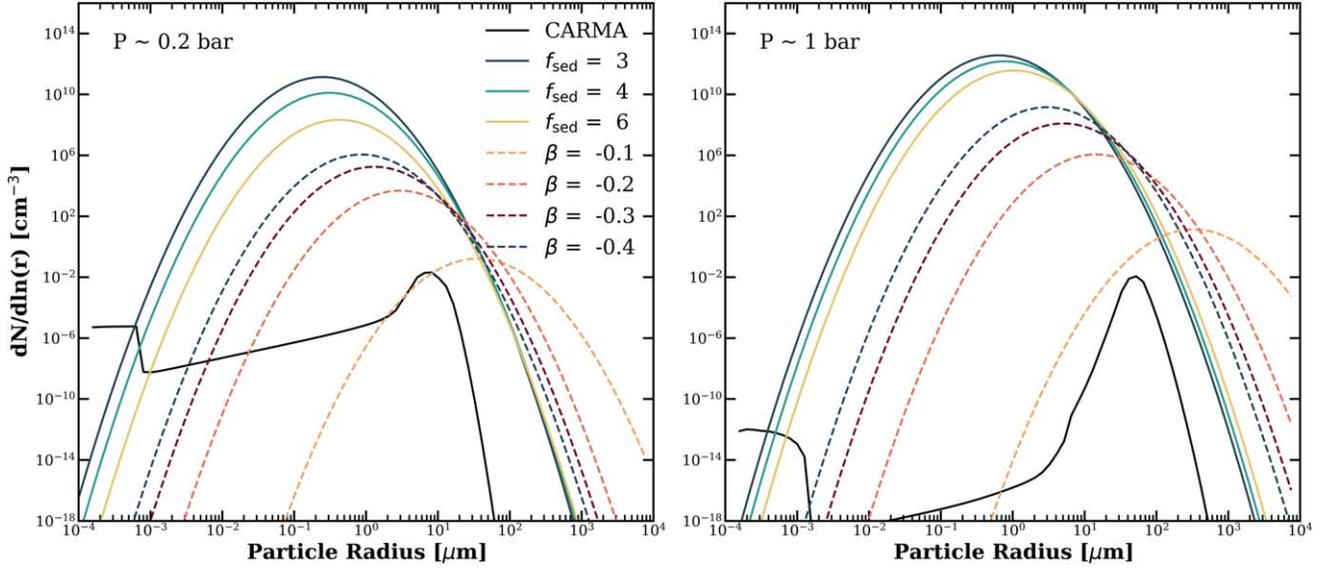

**Figure 15.** Water-ice cloud particle size distribution for a brown dwarf with $T_{eff} = 200$ K and $\log(g) = 4.5$. The distributions of particles are shown at two different layers of the atmosphere with $P \sim 0.2$ bar (left) and 1 bar (right), corresponding to the upper region of the cloud and the cloud base, respectively. The `EddySed` cloud with three constant $f_{sed}$ values (solid) and four variable $f_{sed}$ cases with different $\beta$ values with $\alpha = 3$ (dashed) are compared to the CARMA distribution (black).

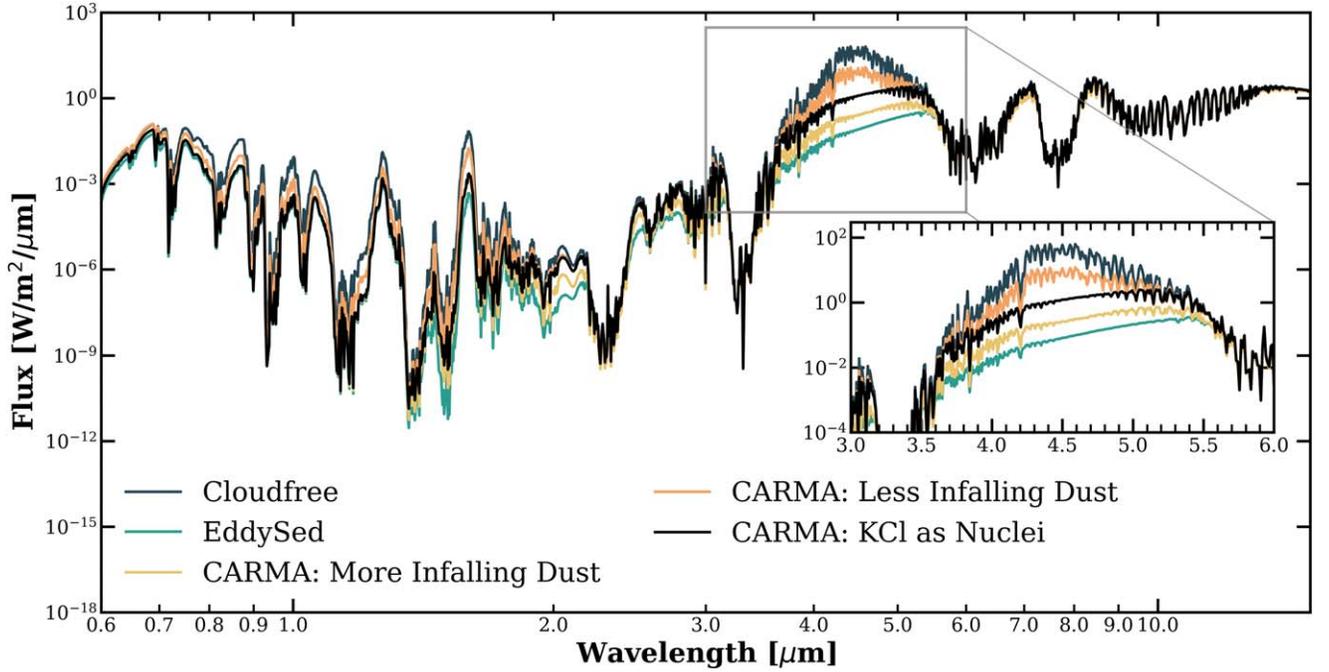

**Figure 16.** Simulated thermal emission spectra from 0.6 to 15 $\mu$m of the CARMA and `EddySed` cloud models and a cloud-free model for a Y dwarf with $T_{eff} = 175$ K, $\log(g) = 4.5$ and $f_{sed} = 6$. The CARMA water-ice clouds include those heterogeneously nucleated on KCl and infalling meteoritic dust. The case with less infalling dust corresponds to a downward flux of $10^{-19}$ g cm$^{-2}$ s$^{-1}$, while more infalling dust corresponds to a flux of $10^{-17}$ g cm$^{-2}$ s$^{-1}$. The inset highlights the region of interest between 3 and 6 $\mu$m.

Tang et al. (2021) for partially cloudy atmospheres at solar metallicity.

We evaluate the necessary latent heat flux needed to significantly impact solar metallicity atmospheres by setting the latent heat flux at every level where water clouds are forming to a constant value. We find that latent heat needs to contribute at least 10 W m$^{-2}$ to impact the P-T profile (Figure 18). Even at 1 W m$^{-2}$, the P-T profile follows that of our fiducial model. For a set of test cases at 200 K, $\log(g) = 4.5$ with supersolar metallicities, the latent heat flux still has a minor impact on the P-T profile in the region where water clouds are forming. Based on a simple extrapolation of our latent heat flux for objects with solar metallicity, we would begin to see the latent heat impacts of water-ice cloud formation on the thermal structure for objects with [M/H] > 1.5. This estimate will vary based on the actual amount of cloud condensation.

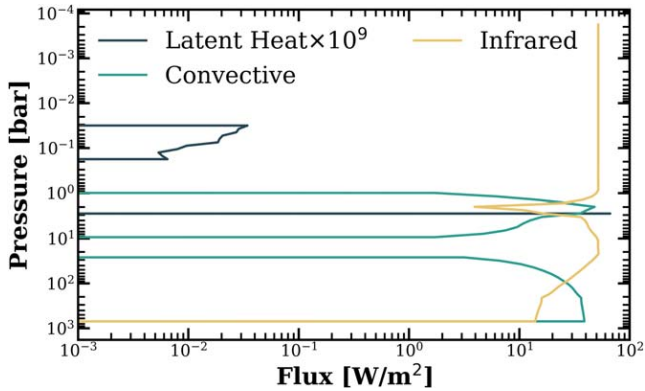

**Figure 17.** Contributions to the total flux in the atmospheric profile of a $T_{eff} = 175$ K, $\log(g) = 4.5$, and $f_{sed} = 6$ brown dwarf. The latent heat flux is scaled up by a factor of $10^9$, due to its negligible contribution, making it visually comparable to the convective and infrared fluxes. The latent heat flux contribution in the radiative zones is significantly lower than the convective and infrared flux contributions.

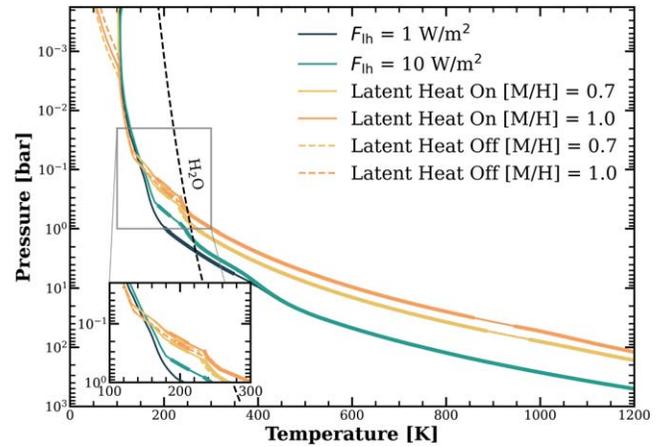

**Figure 18.** Pressure–temperature profiles for a fully cloudy brown dwarf with $T_{eff} = 200$ K and $\log(g) = 4.5$ (cgs) with $f_{sed} = 6$. Two models assume a solar metallicity with the latent heat flux set to 1 and 10 W m$^{-2}$. The other thermal structures, for cases at supersolar metallicity, show the effect of latent heat release in the moist adiabat model (solid) in comparison to those without heat release following a dry adiabat (dashed). The condensation curve of water ice is shown in the black dashed line. The thicker regions of the profile indicate where the atmosphere is convective.

## 4. Discussion

### 4.1. Model Convergence

The time-stepping framework introduced by Robinson & Marley (2014) and extended by Tang et al. (2021) provides a smoothing effect on the P-T profiles in the upper atmosphere and allows for the consideration of radiative cloud feedback. However, it encounters convergence challenges for models with P-T profiles close to the water condensation curve, such as those with an effective temperature of 250 K in a fully cloudy atmosphere as seen in Figure 2. Careful selection of initial radiative–convective boundaries, substep increments, and the initial P-T profile guess greatly impacts the results. We recommend starting with a clear atmospheric P-T profile (e.g., Sonora bobcat models Marley et al. 2021) as the initial guess for a cloudy model, with a radiative–convective boundary guess ∼200 bars.

### 4.2. Vertical Mixing Dynamics

Disequilibrium chemistry is expected within the atmospheres of these cool worlds (Miles et al. 2020; Karalidi et al. 2021; Leggett et al. 2021). Disequilibrium chemistry of molecular species like $NH_3/N_2$, $CO/CH_4$, and possibly $PH_3$ and $CO_2$ can change their abundances in the regions where $H_2O$ clouds form. It is dictated by the balance of the chemical and vertical mixing timescales. The mixing timescale is determined based on the eddy diffusion coefficient $K_{zz}$. In particular for cold substellar objects like those in our work, the chemical timescale will become much longer than the mixing timescale. Observations of disequilibrium chemistry can be used to tune the strength of vertical mixing in the atmosphere.

Mukherjee et al. (2022) investigated the influence of different $K_{zz}$ in cloudless atmospheres, indicating the necessity for an enhanced treatment of this parameter. Currently, $K_{zz}$ in our models is determined using mixing length theory and so is not well constrained in the radiative zone. Mukherjee et al. (2022) calculated $K_{zz}$ in the radiative zone based on the P-T profile as prescribed in Moses et al. (2022) and found that the quench levels of different molecules and the object's mid-infrared spectra are highly sensitive to $K_{zz}$.

We do not use this prescription in this work and do not consider disequilibrium chemistry. Since $K_{zz}$ controls the strength of the vertical mixing in the atmosphere, it will also have a large impact on the cloud morphology. Stronger vertical mixing would upwell more cloud particles and could create more vertically extended clouds. Subsequent investigations will be needed to assess the ramifications on water cloud morphology when considering a P-T profile dependent $K_{zz}$.

### 4.3. Ammonia Clouds

While the colder objects in our grid of models are within the regime where ammonia can begin to condense, we do not include ammonia clouds in our models, since our study focuses on water cloud morphology. The P-T profile of the coldest brown dwarfs in our grid could intersect the ammonia condensation curve in the uppermost regions of the atmosphere. However, due to the low partial pressure of $NH_3$ there, any resulting opacity would be relatively insignificant compared to the extensive water cloud deck situated below. On the other hand, for colder giant planets, ammonia clouds will be a large source of opacity, as seen on Jupiter. This additional opacity will change the observed spectral features and should be included in future models of cold substellar objects.

### 4.4. Variability

Variability has been observed in numerous brown dwarfs, which has typically been interpreted as the heterogeneous distribution of clouds in their atmospheres (Metchev et al. 2015; Cushing et al. 2016; Leggett et al. 2016; Apai et al. 2017; Vos et al. 2022, 2023). Evaluating the dynamics and spatial heterogeneity of water clouds requires 3D general circulation models (GCMs). Alternatively, one-dimensional models can incorporate inhomogeneity through the introduction of patchy clouds, an approach that has been explored in previous studies such as Morley et al. (2014a). Future models combining patchy clouds with a time-stepping framework hold the potential to more realistically portray cloud evolution.

## 5. Conclusions

As we transition into an era of high-resolution observations of ultra-cool substellar objects, it becomes imperative to improve the accuracy of the cloud models we use. This is especially critical for objects with effective temperatures below 400 K, where water clouds are poised to be a significant source of opacity. Here, we introduce new prescriptions for parameterized water cloud models derived from microphysical models, which we propose as guiding principles for forthcoming water cloud models in atmospheres dominated by H/He. We investigate the key components of microphysical clouds using a grid of Y-dwarf models with $T_{\rm eff} = 150–400$ K and $\log(g) = 3.5–5.0$. Throughout this study, we refine the previous treatment of water clouds within the `EddySed` framework, integrating microphysically informed cloud properties and morphology. Our main results include the following:

1. Supersaturation is maintained in the CARMA models, especially for objects warmer than 250 K where the P-T profile of the upper atmosphere sits closer to the water condensation curve, significantly reducing the total amount of water-ice clouds condensing in the atmosphere. The total column mass of the cloud can be matched by `EddySed` by including supersaturation after condensation by a factor of 10 or by removing 90% of the cloud condensate mass mixing ratio. Neither of these approximations produce thermal emission spectra matching those of the CARMA models.

2. Water latent heat release has no significant effect on the water-ice clouds and P-T profiles for objects with solar metallicity atmospheres.

3. A vertically variable $f_{\rm sed}$ can allow the optical depth profile to better match those of the microphysics models. Variable $f_{\rm sed}$ parameters with $\alpha$ values of 3–6 and a $\beta$ value between $-0.1$ and $-1$ generate the best-fit models for our grid. For more vertically extended clouds, two $\beta$ values are required for the best fit.

4. Including subcloud opacity in the `EddySed` models with a ~4% falloff rate from the cloud base optical depth does not significantly aid in reproducing observables. The selected `EddySed` cloud base treatment allows for tracking the microphysical cloud morphology, although its impact is secondary compared to the variable $f_{\rm sed}$, which drives the peak optical depth at the cloud base.

5. Including microphysical particle growth and loss rates in `EddySed` provides an upper limit on the timescales of water-ice cloud formation. This allows for a more physical cloud evolution with the consideration of radiative cloud feedback.

With upcoming observations of ultra-cool objects like Y dwarfs and temperate giant planets by JWST, generating atmospheric models with more accurate water clouds will be essential. Future models will also need to include disequilibrium chemistry, better constraints on $K_{zz}$, patchy clouds for one-dimensional variability studies, and dynamical effects from three-dimensional models. Furthermore, for the coldest objects, it will be imperative to account for additional condensable species, such as $NH_3$.

## Acknowledgments

J.M. acknowledges support from the National Science Foundation Graduate Research Fellowship Program under grant No. DGE 2137420. C.V.M. acknowledges support from the Alfred P. Sloan Foundation under grant number FG-2021-16592. C.V.M. and J.M. acknowledge the National Science Foundation, which supported the work presented here under grant No. 1910969. This material is based on work supported by the National Aeronautics and Space Administration under grant No. 80NSSC21K0650 for the NNH20ZDA001N-ADAP: D.2 program. Support for program JWST-AR-01977.004 was provided by NASA through a grant from the Space Telescope Science Institute, which is operated by the Associations of Universities for Research in Astronomy, Incorporated, under NASA contract NAS5-26555. J.M. would like to thank Mark Marley for his guidance on working with `EGP` and Laura Mayorga for the insightful discussions on model convergence.

*Software*: CARMA (Ackerman et al. 1995), `PICASO` (Batalha et al. 2021), `Jupyter` (Kluyver et al. 2016), NumPy (van der Walt 2011), `SciPy` (Virtanen et al. 2020), and `Matplotlib` (Hunter 2007). Models from this paper are hosted on Zenodo for public use.

## Appendix
## Microphysical Timescale Sensitivity

The microphysical rates we use in this work are directly calculated from CARMA and provide fixed growth and loss rates for each case within our grid. Since we use these rates in `EGP+` to limit cloud growth and loss at each time step, we assess the sensitivity of the steady-state P-T profile to these microphysical timescales. We vary the first term in Equation (10) by factors of 0.01, 0.1, 10, and 100 to explore a wide range of growth and loss rates, thereby generating models that test for differences in the steady-state thermal structure profiles under different microphysical conditions.

As seen in Figure 19, the differences in the steady-state P-T profile are negligible. This indicates that the microphysical timescale does not need to be numerically evaluated for each custom case but can instead rely on the microphysical rates that we have presented in Section 3.4. The microphysical timescales therefore provide a physically informed smoothing effect, enhancing the convergence of the `EGP+` model.

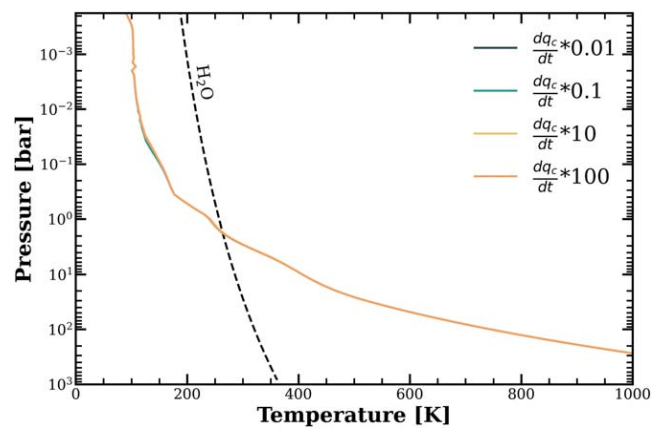

**Figure 19.** Pressure–temperature profiles with various numerical factors for the growth rate for a $T_{\rm eff} = 175$ K, $\log(g) = 4.5$ brown dwarf.

## ORCID iDs

James Mang  https://orcid.org/0000-0001-5864-9599
Caroline V. Morley  https://orcid.org/0000-0002-4404-0456
Tyler D. Robinson  https://orcid.org/0000-0002-3196-414X
Peter Gao  https://orcid.org/0000-0002-8518-9601

## References

Ackerman, A. S., Hobbs, P. V., & Toon, O. B. 1995, JAtS, 52, 1204
Ackerman, A. S., & Marley, M. S. 2001, ApJ, 556, 872
Allard, F., Homeier, D., & Freytag, B. 2011, in ASP Conf. Ser. 448, 16th Cambridge Workshop on Cool Stars, Stellar Systems, and the Sun, ed. C. Johns-Krull, M. K. Browning, & A. A. West (San Francisco, CA: ASP), 91
Apai, D., Karalidi, T., Marley, M. S., et al. 2017, Sci, 357, 683
Batalha, N., & Rooney, C. 2020, natashabatalha/picaso: Release 2.1, v2.1, Zenodo, doi:10.5281/zenodo.4206648
Batalha, N., Rooney, C., & MacDonald, R. 2021, natashabatalha/picaso: Release 2.2, v2.2.0, Zenodo, doi:10.5281/zenodo.5093710
Batalha, N. E., Marley, M. S., Lewis, N. K., & Fortney, J. J. 2019, ApJ, 878, 70
Burrows, A., Ibgui, L., & Hubeny, I. 2008, ApJ, 682, 1277
Cushing, M. C., Hardegree-Ullman, K. K., Trucks, J. L., et al. 2016, ApJ, 823, 152
Cushing, M. C., Kirkpatrick, J. D., Gelino, C. R., et al. 2011, ApJ, 743, 50
Faherty, J. K., Tinney, C. G., Skemer, A., & Monson, A. J. 2014, ApJL, 793, L16
Feng, F., Anglada-Escudé, G., Tuomi, M., et al. 2019, MNRAS, 490, 5002
Fortney, J. J., Lodders, K., Marley, M. S., & Freedman, R. S. 2008, ApJ, 678, 1419
Fortney, J. J., Marley, M. S., & Barnes, J. W. 2007, ApJ, 659, 1661
Fortney, J. J., Marley, M. S., Lodders, K., Saumon, D., & Freedman, R. 2005, ApJL, 627, L69
Fortney, J. J., Visscher, C., Marley, M. S., et al. 2020, AJ, 160, 288
Freedman, R. S., Lustig-Yaeger, J., Fortney, J. J., et al. 2014, ApJS, 214, 25
Freedman, R. S., Marley, M. S., & Lodders, K. 2008, ApJS, 174, 504
Gao, P., Marley, M. S., & Ackerman, A. S. 2018, ApJ, 855, 86
Hubeny, I., & Lanz, T. 1995, ApJ, 439, 875
Hunter, J. D. 2007, CSE, 9, 90
Jacobson, M. Z., Turco, R. P., Jensen, E. J., & Toon, O. B. 1994, AtmEn, 28, 1327
Jensen, E. J., Pfister, L., Ackerman, A. S., Tabazadeh, A., & Toon, O. B. 2001, JGR, 106, 17
Karalidi, T., Marley, M., Fortney, J. J., et al. 2021, ApJ, 923, 269
Kluyver, T., Ragan-Kelley, B., Pérez, F., et al. 2016, in Positioning and Power in Academic Publishing: Players, Agents and Agendas, ed. F. Loizides & B. Schmidt (Amsterdam: IOS Press)
Lacy, B., & Burrows, A. 2023, ApJ, 950, 8
Lacy, B., Shlivko, D., & Burrows, A. 2019, AJ, 157, 132
Leggett, S. K., Cushing, M. C., Hardegree-Ullman, K. K., et al. 2016, ApJ, 830, 141
Leggett, S. K., Morley, C. V., Marley, M. S., & Saumon, D. 2015, ApJ, 799, 37
Leggett, S. K., Tremblin, P., Phillips, M. W., et al. 2021, ApJ, 918, 11
Lodders, K. 1999, ApJ, 519, 793
Lodders, K., & Fegley, B. 2002, Icar, 155, 393
Lodders, K., & Fegley, B. 2006, in Astrophysics Update 2, ed. J. W. Mason (Berlin: Springer), 1
Luhman, K. L. 2014, ApJ, 786, L18
Luhman, K. L., Tremblin, P., Alves de Oliveira, C., et al. 2024, AJ, 167, 5
Lupu, R. E., Marley, M. S., Lewis, N., et al. 2016, AJ, 152, 217
Mang, J., Gao, P., Hood, C. E., et al. 2022, ApJ, 927, 184
Mang, J., Morley, C., Robinson, T., & Gao, P. 2024, Models for Ultra-cool Substellar Objections with Microphysically Informed Water Clouds v1, Zenodo, doi:10.5281/zenodo.13176399
Marley, M. S., Gelino, C., Stephens, D., Lunine, J. I., & Freedman, R. 1999, ApJ, 513, 879
Marley, M. S., & McKay, C. P. 1999, Icar, 138, 268
Marley, M. S., Saumon, D., Cushing, M., et al. 2012, ApJ, 754, 135
Marley, M. S., Saumon, D., & Goldblatt, C. 2010, ApJL, 723, L117
Marley, M. S., Saumon, D., Guillot, T., et al. 1996, Sci, 272, 1919
Marley, M. S., Saumon, D., Visscher, C., et al. 2021, ApJ, 920, 85
Mayorga, L. C., Robinson, T. D., Marley, M. S., May, E. M., & Stevenson, K. B. 2021, ApJ, 915, 41
McKay, C. P., Pollack, J. B., & Courtin, R. 1989, Icar, 80, 23
Metchev, S. A., Heinze, A., Apai, D., et al. 2015, ApJ, 799, 154
Miles, B. E., Skemer, A. J. I., Morley, C. V., et al. 2020, AJ, 160, 63
Morley, C. V., Fortney, J. J., Marley, M. S., et al. 2012, ApJ, 756, 172
Morley, C. V., Fortney, J. J., Marley, M. S., et al. 2015, ApJ, 815, 110
Morley, C. V., Knutson, H., Line, M., et al. 2017, AJ, 153, 86
Morley, C. V., Marley, M. S., Fortney, J. J., & Lupu, R. 2014a, ApJL, 789, L14
Morley, C. V., Marley, M. S., Fortney, J. J., et al. 2014b, ApJ, 787, 78
Morley, C. V., Mukherjee, S., Marley, M. S., et al. 2024, arXiv:2402.00758
Morley, C. V., Skemer, A. J., Allers, K. N., et al. 2018, ApJ, 858, 97
Moses, J. I., Tremblin, P., Venot, O., & Miguel, Y. 2022, ExA, 53, 279
Mukherjee, S., Fortney, J. J., Batalha, N. E., et al. 2022, ApJ, 938, 107
Mukherjee, S., Fortney, J. J., Morley, C. V., et al. 2024, ApJ, 963, 73
Phillips, M. W., Tremblin, P., Baraffe, I., et al. 2020, A&A, 637, A38
Rajan, A., Rameau, J., De Rosa, R. J., et al. 2017, AJ, 154, 10
Robinson, T. D., & Marley, M. S. 2014, ApJ, 785, 158
Rooney, C. M., Batalha, N. E., Gao, P., & Marley, M. S. 2022, ApJ, 925, 33
Saumon, D., & Marley, M. S. 2008, ApJ, 689, 1327
Skemer, A. J., Morley, C. V., Allers, K. N., et al. 2016, ApJL, 826, L17
Sudarsky, D., Burrows, A., Hubeny, I., & Li, A. 2005, ApJ, 627, 520
Tang, S.-Y., Robinson, T. D., Marley, M. S., et al. 2021, ApJ, 922, 26
Toon, O. B., Turco, R. P., Westphal, D., Malone, R., & Liu, M. 1988, JAtS, 45, 2123
Turco, R. P., Hamill, P., Toon, O. B., Whitten, R. C., & Kiang, C. S. 1979, ApJ, 36, 699
van der Walt, S., Colbert, S. C., & Varoquaux, G. 2011, CSE, 13, 22
Virtanen, P., Gommers, R., Oliphant, T. E., et al. 2020, NatMe, 17, 261
Visscher, C., Lodders, K., & Fegley, B. J. 2006, ApJ, 648, 1181
Visscher, C., Lodders, K., & Fegley, B. J. 2010, ApJ, 716, 1060
Vos, J. M., Burningham, B., Faherty, J. K., et al. 2023, ApJ, 944, 138
Vos, J. M., Faherty, J. K., Gagné, J., et al. 2022, ApJ, 924, 68
Zhang, Z., Liu, M. C., Marley, M. S., Line, M. R., & Best, W. M. J. 2021, ApJ, 921, 95